# Physics of Chiral Photonic Crystals with Defect

Ashot H. Gevorgyan[1], Koryun B. Oganesyan[2,3*] Edik A. Ayryan[3],
Michal Hnatic[4,5,6], Yuri V. Rostovtsev[7], Gershon Kurizki[8]

[1] Department of Physics, Yerevan State University, Yerevan, Armenia

[2] Alikhanyan National Science Lab, Yerevan Physics Institute, Alikhanyan Br.2, 036, Yerevan, Armenia

[3] LIT, Joint Institute for Nuclear Research, Dubna, Russia

[4] Faculty of Sciences, P. J. Safarik University, Park Angelinum 9, 041 54 Kosice, Slovakia

[5] Institute of Experimental Physics SAS, Watsonova 47, 040 01 Kosice, Slovakia ˇ

[6] BLTP, Joint Institute for Nuclear Research, Dubna, Russia

[7] University of North Texas, Denton, TX, USA

[8] Weizmann Institute of Science, Rehovot, Israel

[*] bsk@yerphi.am

**Abstract**—Specific features of the defect modes of cholesteric liquid crystals (CLCs) with an isotropic defect, as well as their photonic density of states, $Q$ factor, and emission, have been investigated. The effect of the thicknesses of the defect layer and the system as a whole, the position of the defect layer, and the dielectric boundaries on the features of the defect modes have been analyzed.

## 1. INTRODUCTION

Cholesteric liquid crystals (CLCs) are the most widespread representatives of 1D chiral photonic crystals (CPCs) due to the possibility of spontaneous selforganization of their periodic structure and controllability of their photonic band gap (PBG) in a wide frequency range. The CLC parameters can be varied by an external electric/magnetic field, temperature gradient, UV radiation, etc. CLCs have also some other surprising optical properties. The main difference between CLCs and conventional photonic crystals is that in the former a PBG exists only for light with one circular polarization (at normal incidence of light), which coincides with the chirality sign of the medium. In these crystals Bragg reflection occurs in the spectral range from the wavelength $\lambda_1 = pn_o$ to $\lambda_2 = pn_e$ ($p$ is the helical pitch and $n_o = \sqrt{\varepsilon_1}$ and $n_e = \sqrt{\varepsilon_2}$ are respectively, the local ordinary and extraordinary refractive indices). Light with opposite circular polarization does not undergo diffractive reflection. Currently, these media are of great interest because of the possibility of low-threshold lasing at the edges of their PBG (predicted by Dowling et al. [1] and experimentally confirmed by Koop et al. [2]).

There is increasingly interest in study of CLC with defects, see e.g. [3-87].

In this paper, we report the results of the further study of the specific features of defect modes in CLCs with an isotropic defect.

## 2. METHOD OF ANALYSIS

The transmission of a plane-polarized wave through a CLC layer with an isotropic defect will be analyzed by the modified Ambartsumyan method of adding layers [9, 13]. A CLC with an isotropic defect can be considered as a three-layer system composed of two CLC layers (CLC(1) and CLC(2)) with an isotropic dielectric layer (IDL) between. Let a wave with a complex amplitude $\mathbf{E}_i$, normally fall on this system. We will denote the complex amplitudes of the reflected and transmitted fields as $\mathbf{E}_r$ and $\mathbf{E}_t$, respectively, and expand them in the basic $p$ and $s$ polarizations: $\mathbf{E}_{i,r,t} = E^p_{i,r,t}\mathbf{n}_p + E^s_{i,r,t}\mathbf{n}_s = \begin{pmatrix} E^p_{i,r,t} \\ E^s_{i,r,t} \end{pmatrix}$ ($\mathbf{n}_p$ and $\mathbf{n}_s$ are the unit vectors of the $p$ and $s$ polarizations, respectively). Then, we can write the solution as

$$\mathbf{E}_r = \hat{R}\mathbf{E}_i, \quad \mathbf{E}_t = \hat{T}\mathbf{E}_i \tag{1}$$

where $\hat{R}$ and $\hat{T}$ are the 2 × 2 reflection and transmission matrices of the system, respectively. According to [9, 13], for a system composed of two adjacent layers ($A$ on the left and $B$ on the right), the reflection and transmission matrices of the $A + B$ system ($A + B$" $RA + B$ and $TA + B$, respectively) are determined in terms of the similar matrices of the layers by the matrix equations

$$\hat{R}_{A+B} = \hat{R}_A + \tilde{\hat{T}}_A \hat{R}_B \left(\hat{I} - \tilde{\hat{R}}_A \hat{R}_B\right)^{-1} \hat{T}_A, \quad \hat{T}_{A+B} = \hat{T}_B \left(\hat{I} - \tilde{\hat{R}}_A \hat{R}_B\right)^{-1} \hat{T}_A, \tag{2}$$

where $\hat{I}$ is the unit matrix and tilde indicates the corresponding reflection and transmission matrices for the backward propagation. To obtain the reflection and transmission matrices of the system under consideration, we will use formula (2) first to match the left side of the isotropic layer with the CLC(2) layer and then to match the left side of the system obtained with the CLC(1) layer.

The results of the analysis are reported below.

## 3. RESULTS AND DISCUSSION

Let us consider the normal incidence of light on the CLC(1)–IDL–CLC(2) multilayer system from the left side. Calculations were performed for a CLC layer with $n_o = \sqrt{\varepsilon_1} = 1.4639$ and $n_e = \sqrt{\varepsilon_2} = 1.5133$ (the CLC composition is cholesteryl nonanoate : cholesteryl chloride : cholesteryl acetate = 20 : 15 : 6), which has a helical pitch $p = 0.42$ μm in the optical range at room temperature (24°C). The CLC spiral is right-handed; therefore, there is a PBG for right-handed circularly polarized light incident on the defect-free CLC layer and no such a band for left-handed circularly polarized light.

The presence of a thin defect in the CLC structure is known to initiate a defect mode in the PBG. This mode manifests itself as a dip in the reflection spectrum for right-handed circularly polarized light (diffracting circular polarization) and as a peak in the reflection spectrum for left-handed circularly polarized light (nondiffracting circular polarization). The defect mode is of either donor or acceptor type, depending on the optical thickness of the defect layer: the defect-mode wavelength increases from the minimum to the maximum of the band gap with an increase in the optical thickness of the defect; in this case, two defect modes arise near both gap edges. With a further increase in the optical thickness of the defect layer, the long_wavelength mode is rejected, and the short-wavelength mode becomes red-shifted [3].

In the case of a thick defect layer ($d^d \gg \lambda$) the number of defect modes increases. The frequency position of these modes and the number of excited modes can be varied by changing the thickness of the defect layer.

In the case of an anisotropic defect, due to anisropy, an additional phase difference arises, which results in some important features with an increase in the defect thickness. In particular, the half-width of the defect mode becomes dependent on the thickness of the defect layer and, for example, at $d^d \sim \lambda/2(n_e^d - n_o^d)$ (i.e., when the defect is a half-wave plate), the total (nonselective with respect to polarization) reflection occurs in the PBG [9, 18].

In the case of an isotropic defect the half-width of the defect mode changes only slightly with a change in the defect thickness. Under certain conditions, this circumstance becomes an important advantage.

Indeed, in the case of an anisotropic defect, the efficient accumulation of light, a high $Q$ factor for the defect mode, and low-threshold lasing can only be obtained with a very thin planar defect layer, which can hardly be obtained in practice.

In the case of an isotropic defect all the aforesaid can be implemented with relatively thick defect layers (see below).

As was noted in the introduction, CLCs enriched in laser dyes (resonant atoms) can be used to design feedback lasers and lasers without mirrors (under certain conditions). In amplifying media (in particular, CLCs enriched in fluorescent guest molecules so as the fluorescence peak lies in the PBG or includes the latter), the PBG significantly affects the emission spectrum. Within the PBG, a wave decays and its amplitude exponentially decreases (evanescent wave), due to which the spontaneous emission is suppressed.

This can be explained as follows. The PDOS tends to zero and, since the spontaneous emission intensity is proportional to the PDOS, this intensity also tends to zero. The spontaneous emission lifetime τs sharply increases at the PBG edges (τs decreases with oscillations beyond the PBG), and the stimulated emission increases significantly. It is well known that PDOS has peaks at the PBG edges in the absence of a defect. In the presence of a defect a high PDOS peak is observed for the defect mode and, simultaneously, the PDOS is partially suppressed at the PBG edges.

Let the CLC layer with an isotropic defect be enriched in dye molecules. In the presence of a pump wave this system is amplifying; i.e., can be considered as a planar cavity with an active element. The presence of dye molecules in the system changes its local refractive indices. In this case, the effective imaginary parts of the effective local refractive indices of both CPC ($n_{1,2}^{''}$) and the isotropic defect ($n^{''d}$) are negative ($n_{1,2} = n_{1,2}^{'} + i n_{1,2}^{''}$ and $n^d = n^{'d} + i n^{''d}$). In the presence of absorption (in this case, the imaginary parts $n_{1,2}^{''}$ and $n^{''d}$ of the local refractive indices of the CLC and isotropic defect are positive), the quantity $A = 1 - (R + T)$, which characterizes the light energy absorbed in the system ($R$ and $T$ are the power reflectance and transmittance, respectively), is positive and less than unity. In an amplifying medium, $A$ is negative; hence, the emission from the system will be characterized by the magnitude $|A|$. Let us assume that the $n''_{1,2}$ and $n''^{d}$ values are negative (i.e., the system is amplifying). If $|n''_{1,2}|$ and $|n''^{d}|$ are rather small, the waves emerging from the system would exist only in the presence of an external wave incident on the system; in this case, the value $|A|$ characterizes the amplifying system. However, if the imaginary parts $|n''_{1,2}|$ and $|n''^{d}|$ reach certain values, $R$ and $T$ sharply change at certain frequencies (depending on the parameters of the system) and tend to infinity; thus, the amplitudes of the waves emerging from the system can be nonzero even at zero amplitude of the incident wave.

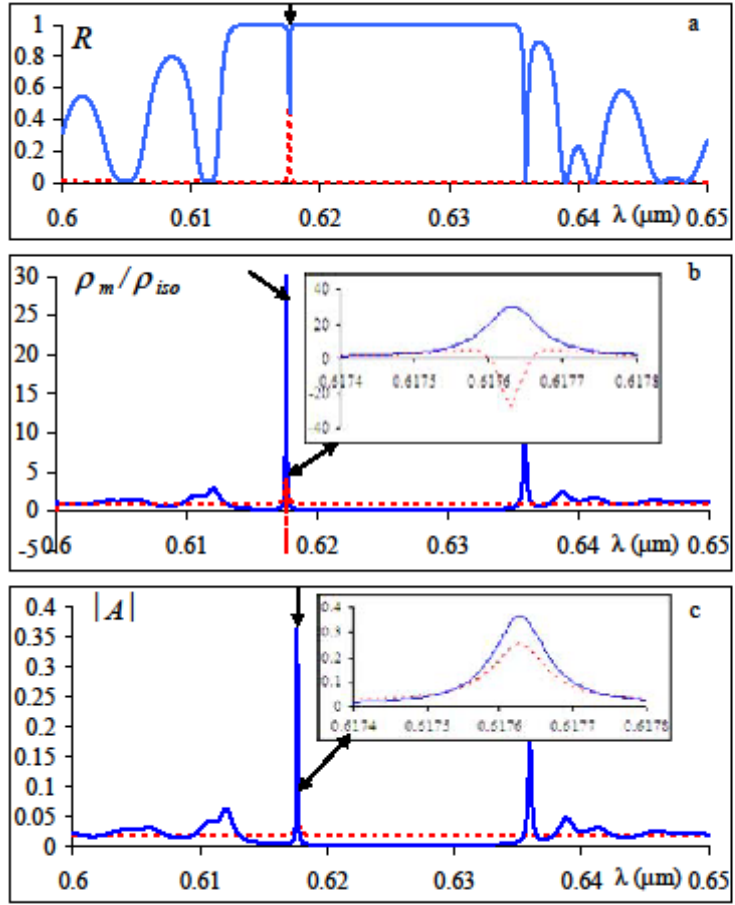

**Fig. 1.** Spectra of (a) reflectance, (b) relative PDOS $\rho m/\rho iso$, and (c) emission intensity $|A|$ for a CLC layer with an isotropic defect in the midplane. Light incident on system is right-handed (solid line) or left-handed (dashed line) circularly polarized. CLC helix is right-handed. CLC parameters are as follows: $\varepsilon 1 = 2.29-0.0001i$, $\varepsilon 2 = 2.143-0.0001i$, $Sd = 0$, helical pitch $\sigma = 0.42$ μm, CLC thickness $d = 70\sigma$, defect-layer thickness $dd = 1.86$ μm, defect-layer refractive index $nd = 1.7$, and refractive index of the medium $n0 = \sqrt{(\varepsilon_1 + \varepsilon_2)/2}$ around system.

Obviously, in this case the amplitudes of the reflected and transmitted waves cannot be determined by solving the linear problem, and the nonlinear problem must be solved. However, now, one can use the condition of nonzero solutions for the amplitudes of the reflected and transmitted waves at zero amplitude of the incident wave to determine the so called laser-mode frequencies and the corresponding gains, i.e., the minimum threshold gain at which lasing occurs (see similar considerations for a periodic layered structure and a chiral periodic medium in [87-90]).

As was shown in [89, 90], for CLCs with an ideal periodic structure at low amplification and small values of the parameter $d\text{Im}k$ ($d$ is the CPC thickness and $k$ is the wavenumber of the diffracting mode in the rotating coordinate system), the aforementioned condition can be written analytically. It is generally (particularly in our case) difficult to derive analytical expressions for the laser-mode frequencies and the corresponding gains. However, one can determine the frequencies of possible laser modes and the corresponding threshold gains from the sharp peaks of $|A|$. Note that the features of the absorption (emission) of ideal CLC were numerically analyzed in [91,92].

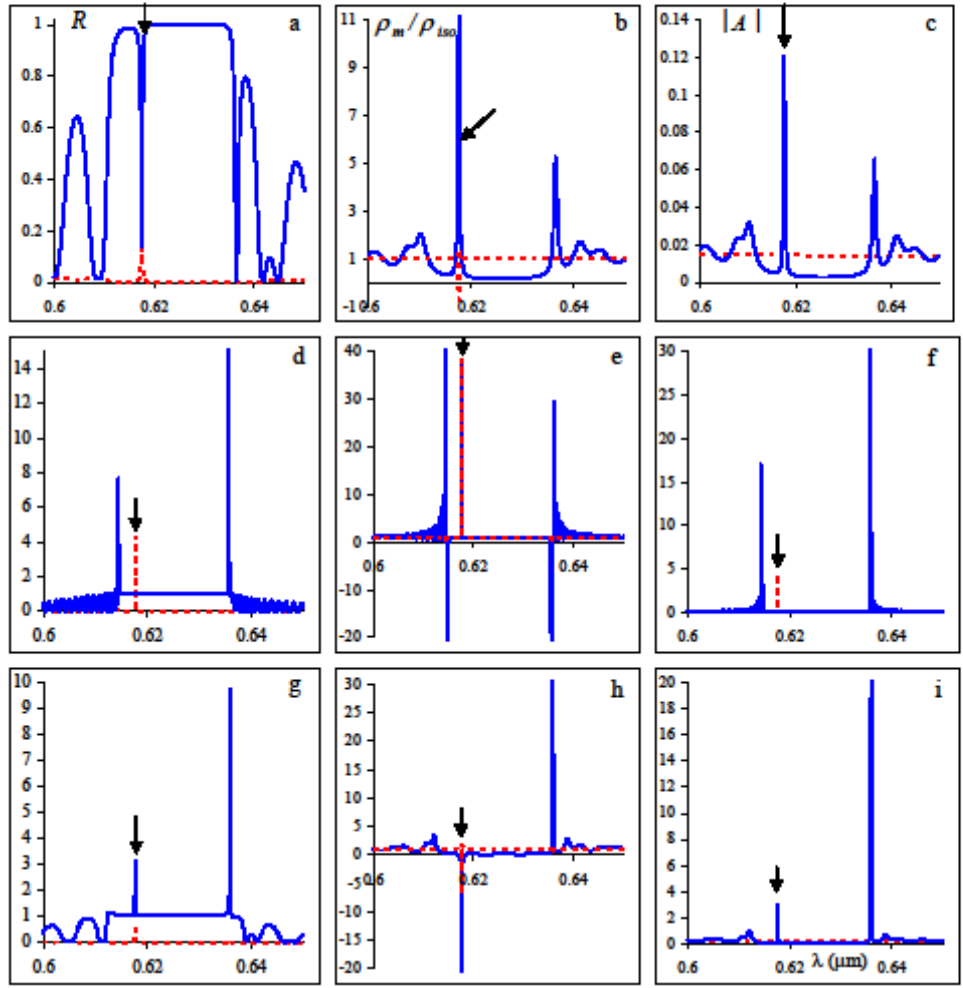

**Fig. 2.** Spectra of reflectance $R$, relative PDOS $\rho m/\rho iso$, and emission intensity $|A|$ for different CLC-layer thicknesses and gains. In the first row $d = 50\sigma$ and $\varepsilon_1'' = \varepsilon_2'' = -0.0001$; in the second row $d = 600\sigma$ and $\varepsilon_1'' = \varepsilon_2'' = -0.0001$; and, in the third row, $d = 70\sigma$ and $\varepsilon_1'' = \varepsilon_2'' = -0.001$. Parameters of system and curves are same as in Fig. 1.

Furthermore, we will characterize the degree of order of the dipole moments of the transitions in guest molecules by the order parameter $S_d$. This parameter is determined in terms of the mean of $\cos\vartheta$ as $S_d = \frac{3}{2}\langle\cos\vartheta\rangle - \frac{1}{2}$, where $\vartheta$ is the angle between the local direction of the CPC optical axis and the dipole moment of the transition in guest molecules. The maximally possible order parameter $S_d = 1$ corresponds to the ideal orientation of the dipole-transition moments (i.e., parallel to the local direction of the optical axis). The value $Sd = 0$ corresponds to isotropic distribution of orientations, and the minimum value $Sd = -0.5$ corresponds to the isotropic distribution of dipole-transition moments in a plane perpendicular to the local optical axis. Within the linear-optics approximation the relations obtained describe both the amplification and generation regimes.

Figure 1 shows the wavelength dependences of (a) the reflectance $R$, (b) the relative PDOS $\rho_m/\rho_{iso}$, and (c) the emission intensity in the presence of an isotropic defect in the midplane of the CLC layer. The light incident on the system is right-handed (solid line) and left-handed

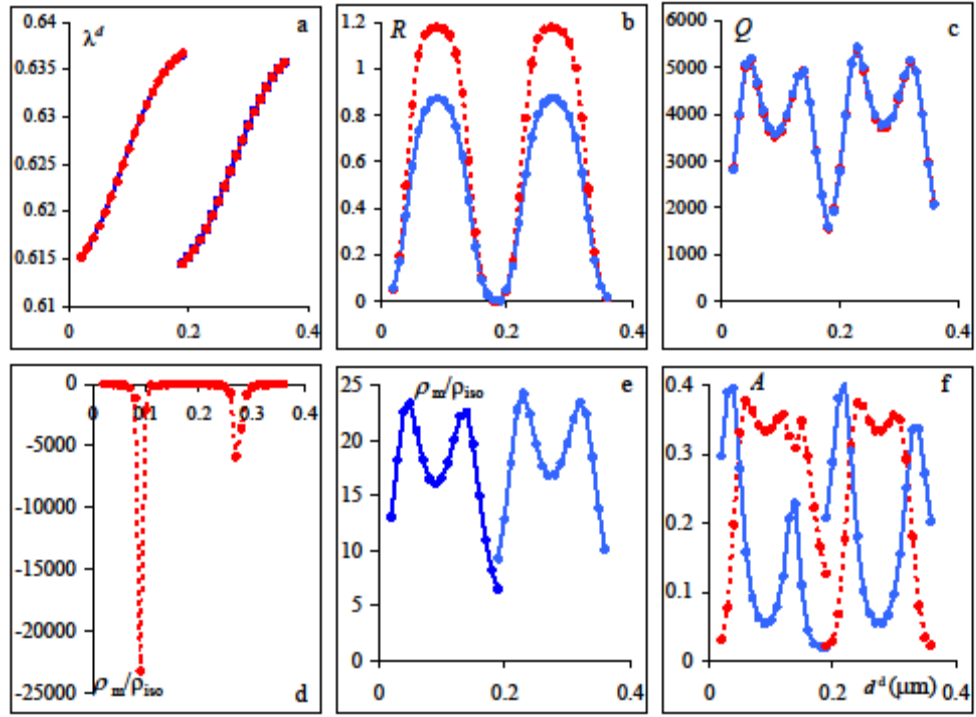

**Fig. 3.** Dependences of (a) defect-mode wavelengths $\lambda^d$ and (b) reflectances $R$, (c) $Q$ factors, (d, e) relative PDOSs $\rho_m/\rho_{iso}$, and (f) emission intensities $|A|$ at defect modes on defect-layer thickness $d^d$. Parameters of system and curves are same as in Fig. 1. (dashed line) circularly polarized.

As can be seen in Fig. 1, the presence of a defect in the CLC structure leads to the occurrence of defect modes in the PBG (indicated by an arrow). In the PBG (which is in the range of $\lambda = 0.6148$–$0.6356$ μm for the given parameters of the problem) one can observe a resonant decrease in emission (parameter $|A|$) for right-handed circularly polarized light, similar to the diffractive suppression of absorption. The emission at the PBG edges is anomalously intense (by an analogy with the effect of anomalous absorption near the PBG edges).

For the defect mode with the central wavelength λ = 0.61763 μm (i.e., near the short-wavelength PBG boundary), ρm/ρiso has a sharp peak for right-handed circularly polarized light and a sharp dip for left-handed circularly polarized light. For the defect mode, the value $|A|$ has a sharp peak for light of both circular polarizations, and the emission of light of arbitrary polarization is anomalously high. This means that low-threshold lasing may occur on this mode.

Furthermore, a comparison of these results with similar data for the structure without a defect shows that the presence of a defect somewhat suppresses the emission at the PBG edges. Note also that the reflectance $R$, the relative PDOS $\rho_m/\rho_{iso}$, and the emission intensity $|A|$ for left_handed circularly polarized light exhibit no features at the PBG edges because we consider the case of a minimum effect of dielectric bound-aries (i.e., $n_0 \neq \sqrt{(\varepsilon_1+\varepsilon_2)/2}$ ). When $n_0 \neq \sqrt{(\varepsilon_1+\varepsilon_2)/2}$, the spectra of the reflectance $R$, the relative PDOS $\rho_m/\rho_{iso}$, and the emission intensity also contain oscillations near the PBG edges for left-handed circularly polarized light.

The effect of the CLC layer thickness on the defect modes was investigated in [17, 22]. It was shown that a right-handed circularly-polarized defect mode is strongly excited when the CLC layer is thin, whereas the defect mode with left-handed circular polarization is strongly excited in the case of a thick layer. At intermediate values of the CLC layer thickness defect modes with both circular polarizations are excited.

Similar results were obtained in [6, 93] for a defect caused by the helical-phase jump at the interface between two CLC layers. Later the effect of change in the gain ($\varepsilon''_{1,2}$) on the emission from system was investigated in [21, 22]. It was shown that the defect mode emission is suppressed with an increase in Im($\varepsilon''_{1,2}$), whereas the emission at the wavelengths corresponding to the first minima of the reflectance (beyond the PBG but near its edges) is enhanced. With a further increase in $\varepsilon''_{1,2}$, the emission peaks shift to the high order reflectance minima.

We investigated the effect of a change in the CLC layer thickness and gain on the reflection features, the relative PDOS ρm/ρiso, and the emission intensity . The spectra of the reflectance $R$, the relative PDOS $\rho_m/\rho_{iso}$, and the gain are shown in Fig. 2. The defect is in the midplane of the system. As can be seen in Figs. 2b and 2c, the relative PDOS and emission intensity have local peaks both at the defect mode and near the PBG edges at the reflectance minima; the relative PDOS at the defect mode exceeds greatly its value for the reflectance minima at the PBG edges.

Our analysis for right handed circularly polarized light showed that an increase in the CLC layer thickness causes again local peaks of the relative PDOS both at the defect mode and at the reflectance minima near the PBG edges; however, now the relative PDOS value at the first reflectance minimum (near the shortwavelength PBG edge) exceeds that at the defect mode (Fig. 2e). For left-handed circularly polarized light the relative PDOS is maximum at the defect mode.

A further increase in the CLC layer thickness suppresses the emission at the defect mode and gives rise to peaks in $|A|$ at the PBG edge frequencies (for right-handed circularly polarized light). For left_handed circularly polarized light, the emission is maximum at the defect mode (Fig. 2f). With an increase in the gain the maximum of the relative PDOS $\rho_m/\rho_{iso}$ for right handed circularly polarized light is replaced with a minimum (Fig. 2h). With an increase in , the emission intensity at the local maximum of the defect mode is much lower than that at the first reflection minimum near the long-wavelength PBG edge (Fig. 2i).

Let us now consider how a change in the defect-layer thickness affects the defect-mode features. When the defect layer has a finite thickness, the system under study is a microcavity, the main characteristic of which is the $Q$ factor. Therefore, we also investigated the dependence of the $Q$ factor ($Q = \lambda/\Delta\lambda$, where $\Delta\lambda$, is the defect mode half width) on the defect thickness.

Figure 3 shows the dependences of (a) the defect-mode wavelengths $\lambda^d$, (b) the reflectances $R$, (c) the $Q$ factors, (d, e) the relative PDOSs $\rho_m/\rho_{iso}$, and (f) the emission intensities $|A|$ (at the defect modes) on the defect-layer thickness $d^d$ for right-handed (solid line) and left-handed (dashed line) circularly polarized light incident on the system.

Note that the dependences of these parameters on the layer thickness are cyclic, and they are varied in nearly the same ranges in each cycle, which does not hold true for an anisotropic defect [9]. In the latter case, the defect line half width increases from cycle to cycle (whereas the $Q$ factor decreases) and reaches the PBG width when the defect is a half-wave plate [9].

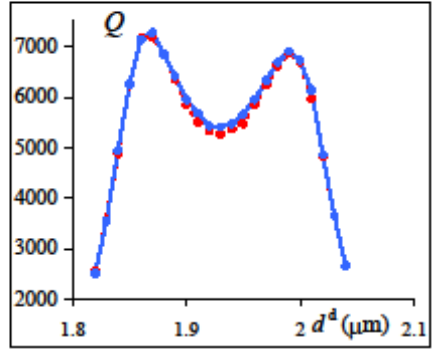

**Fig. 4.** Dependence of defect-mode $Q$ factors on defect layer thickness $d^d$. Parameters of system and curves are same as in Fig. 1.

In the case under consideration, vice versa, the $Q$ factor increases from cycle to cycle, which may also ensure low-threshold lasing for thick defect layers. For comparison, Fig. 4 shows the dependence of the $Q$ factor on the defect thickness for other ranges of variation in this thickness. Similarly, the relative PDOS ρm/ρiso (for the right-handed circular polarization) and the emission intensity $|A|$ somewhat increase from cycle to cycle.

## 4. CONCLUSIONS

We investigated the specific features of the defect modes of CLCs with an isotropic defect (specifically, the features of the reflection and the change in the wavelengths of the defect modes, relative photonic densities, emission intensities, and $Q$ factor) with a change in the defect-layer thickness, the position of the defect in the system, the CLC layer thickness, and the refractive index of the medium around the system. It was shown that, unlike the case of an anisotropic defect, in a system with an isotropic defect the half-width of the defect mode changes only slightly with a change in the defect thickness. This circumstance allows one to efficiently accumulate light and obtain low-threshold defect-mode lasing with a relatively thick defect layer, a situation that can easily be implemented experimentally.